\documentclass[journal,comsoc]{IEEEtran}
\usepackage[T1]{fontenc}

\ifCLASSINFOpdf
\else
\fi
\usepackage{amssymb,amsbsy,epsfig,float,bm}
\usepackage{graphicx}
\usepackage{dblfloatfix}
\usepackage{subfigure}
\usepackage{balance}
\usepackage{algorithm}
\usepackage[noend]{algpseudocode}
\usepackage{wrapfig,tabularx}
\usepackage[hyphens]{url}
\usepackage[hidelinks]{hyperref}
\hypersetup{breaklinks=true}
\usepackage{algorithm}
\usepackage[noend]{algpseudocode}
\usepackage{subfigure}
\usepackage{float}
\restylefloat{table}
\usepackage{xcolor}
\definecolor{LightCyan}{rgb}{0.88,1,1}
\usepackage{color, colortbl}
\def\BibTeX{{\rm B\kern-.05em{\sc i\kern-.025em b}\kern-.08em
    T\kern-.1667em\lower.7ex\hbox{E}\kern-.125emX}}
\bstctlcite{IEEEexample:BSTcontrol}

\usepackage{amsmath}
\usepackage[cmintegrals]{newtxmath}
\begin{document}
%
\title{FairNet: A Measurement Framework for Traffic Discrimination Detection on the Internet}
%
%
%


\ifCLASSOPTIONpeerreview
\author{Vinod~Khandkar,
        and~Manjesh~Hanawal
}
\fi

\author{\IEEEauthorblockN{Vinod S. Khandkar and Manjesh K. Hanawal} \\
	\IEEEauthorblockA{\textit{IEOR, Indian Institute of Technology Bombay, India} \\
		\{vinod.khandkar, mhanawal\}@iitb.ac.in}
}

\maketitle

\begin{abstract}
Network neutrality is related to the non-discriminatory treatment of packets on the Internet. Any deliberate discrimination of traffic of one application while favoring others violates the principle of neutrality. Many countries have enforced laws against such discrimination. To enforce such laws, one requires tools to detect any net neutrality violations. However, detecting such violations is challenging as it is hard to separate any degradation in quality due to natural network effects and selective degradation. Also, legitimate traffic management and deliberate discrimination methods can be technically the same, making it further challenging to distinguish them. 

We developed an end-to-end measurement framework named FairNet to detect discrimination of traffic. It compares the performance of similar services. Our focus is on HTTPS streaming services which constitute a predominant portion of the Internet traffic. The effect of confounding factors (congestion, traffic management policy, dynamic rate adaptation) is made `similar' on the test services to ensure a fair comparison. FairNet framework uses a ``replay server'' and user-client that exchanges correctly identifiable traffic streams over the Internet. The Server Name Indication (SNI) field in the TLS handshake, which goes in plaintext, ensures that the traffic from the replay server appears to network middle-boxes as that coming from its actual server.  We validated that appropriate SNIs results in the correct classification of services using a commercial traffic shaper. FairNet uses two novel algorithms based on application-level throughput and connection status to detect traffic discrimination. We also validated the methodology's effectiveness by collecting network logs through mobile apps over the live Internet and analyzing them.
\end{abstract}

\begin{IEEEkeywords}
Net neutrality, Traffic Discrimination, Service Name Indication (SNI), HTTPS traffic, and Traffic shapers
\end{IEEEkeywords}

%
\IEEEpeerreviewmaketitle

\section{Introduction}
\label{sec:intro}
As commercial activities on the Internet have surged, the expectations on the Internet have gradually emerged towards `guaranteed service' rather than  'best-efforts service,' which was its original design goal. This paradigm shift has tempted Internet Service Providers (ISPs) to use various means to maintain high quality for more commercially viable services and degrade quality for other commercially less viable services that include services generated from the competitors. There are growing cases of unjustified service-specific throttling \cite{wehe_res_2019} and site-blocking \cite{rjio-site-blocking} that are very recent. Also, there are regulatory provisions for ISPs to perform ``reasonable" traffic management that could impact the performance of specific applications but increase the overall experience for users \cite{eu_allows_throttling}. Under such a complex ecosystem, regulators opt for net neutrality (NN) as a guideline to distinguish malicious from lawful traffic management practices (TMPs). To enforce any neutrality laws, one has to have a mechanism to identify any deliberated discrimination. This work proposes a method to detect any deliberate traffic discrimination (TD) or NN violations on the Internet.

Our focus in this work is on audio and video media streaming services. These services occupy over $66\%$ of market share with $50\%$ of services using 4K HD videos (requiring around 25 Mbps speed) \cite{int_study}.  These media streaming services use Dynamic Adaptive Streaming over HTTP (DASH) to maintain the required Quality of Service (QoS) and provide enriched user experience. These schemes are proprietary and could differ across the content providers. Moreover, these services now use HTTPS protocol which provides the end-to-end encryption of the services and makes it difficult for the ISPs to look into the payload and classify the traffic. 

There have been many tools developed for detecting traffic differentiation \cite{nn_survey_2018}. Most of the tools use client-server architecture in which traffic of services to be tested for discrimination is replayed from a dedicated server— the contents from services of interest are copied from the original server and stored in the server. Thus the server acts as a proxy of the original server. To check whether a service is subjected to discrimination, the client accesses that service from the server and compares the QoS it receives against reference traffic. The primary hypothesis for detection is that traffic accessed from the server (target flow) is treated from the network middle-boxes as though it is originated from the original server and subjected to intended discrimination. In contrast, the reference flow will not be classified by the middle-boxes and will evade discrimination. Thus, comparing the QoS of the two flows will reveal any discrimination of the service of interest. 

The existing tools suffer many limitations when used to detect discrimination of HTTPS traffic. The primary issue is that the network middle-boxes may not treat the targeted traffic generated from the server in the same way they treat the original server traffic and lead to incorrect classification of services. Mis-classification could be due to the following reasons: 1) servers used for replay may not capture the application server's DASH mechanism. 2) the existing tools use port $80$ to replay encrypted traffic, which is by default treated as un-encrypted traffic and may not detect any useful signatures. 3) The reference traffic used for comparisons with target traffic may evade the classification but may be treated as some default class and receive a different QoS from the intended service. Thus comparisons of target and reference traffic performance may not lead to correct conclusions.

To overcome these issues, we developed a new framework for measurement named ``FairNet". FairNet also uses a client-server architecture. The server, referred as ``common server" (CS), hosts contents of all the services of interest taken from original servers (by sniffing or direct downloading) and replays them on demand. FairNet overcomes the limitation of the exiting tools as follows: 1) The CS uses a custom DASH so that the traffic is adapted to the actual network environment making the replay traffic appear more genuine. Customized DASH at the CS avoids bandwidth overshoot above available bandwidth and ensures similar data transmission rates for all services under consideration. 2) FairNet sends Server Name Indication (SNI) field in the TLS handshake appropriate for the service. SNI field goes unencrypted in the TLS handshake phase and is directly readable by the middle-boxes. 3) Instead of comparing target traffic with reference traffic, we compare the performance of traffic of service with traffic of another service of a similar type. For example, to check if YouTube is being discriminated against, we compare it to another service like Netflix or Hotstar that has similar service requirements.  

FairNet measurement framework is based on sound principles. Network middle-boxes look for plain-text information in addition to the other signature they could extract from the encrypted payload, e.g., packet size, inter-arrival time. Server Name Indication (SNI) field in the TLS handshake is one such field that goes in plain text and is thus reliably used by the network middle-boxes to classify traffic of different services \cite{trcl_survey}. If one notices any performance degradation of a specific traffic stream (say YouTube), it is hard to pinpoint the actual cause of the degradation. Instead, we can compare the performances of similar traffic streams (say YouTube vs Prime Video or YouTube vs Netflix) to measure the relative performances where deliberate discrimination could become apparent. Thus, it is important to compare services requiring similar QoS for the comparisons to result in meaningful conclusions. Comparison of traffics requiring similar QoS requirements to detect TD is also one of the recommendations by the Body of European Regulators for Electronic Communications (BEREC) \cite{berec_1101} to detect deliberation discrimination on the Internet. 

Our traffic discrimination (TD) detection algorithm compares the throughput performance of a target service with that of another service requiring similar QoS. Use of the CS, DASH, simultaneous access of services, and comparison of similar types of services ensures that the effect of confounding factors on QoS (delay, jitters, throughput) will be largely similar. In the Internet, packets can take different path even if originated from a common source, inducing performance difference. This difference can accentuate if the base rates are different across the service. Our DASH mechanism mitigates this by using the same base rate across all the services. Hence any  Thus, any significant variation in the performance of two traffic streams can be attributed to deliberate TD. 

\begin{itemize}
\item We provide a mechanism to make the traffic replayed from the CS appears similar to that of the original server's traffic by the middle-boxes. FairNet uses the original service's SNI value in the initial TLS handshake message. Due to this parameter setting, the network middle-boxes correctly classify traffic replayed from the CS. We verify using a commercial traffic shaper that all our replayed traffic are classified as intended services (refer Sec.~\ref{sec:css_os_tr_similarity}). 
\item We ensure that the effect of network confounding factors are normalized across all services of similar types simultaneously accessed from the CS. FairNet exchanges traffic between dedicated client and server simultaneously using a customized DASH. This normalizes the network effects such as congestion, network TMPs (refer Sec.~\ref{subsec:perf_compare}), and server-side data transmission mechanisms (refer Sec.~\ref{sec:comm_server}). Thus all services under test are expected to experience similar network effects, and any measured performance can be used for direct comparison to detect any deliberate discrimination.     
\item FairNet uses the novel TD detection algorithm (refer Sec. \ref{sec:tddetectalgo}) based on downloaded file size, connection breaks, and received throughput performance. It directly compares the average throughput within a time slot across services while considering the performance similar to that of the service under test to avoid false detections.
\item  We built an Android and iOS app and a CS hosted on the cloud that implements an end-to-end FiarNet measurement system. The apps are publicly available on Play Store and App Store(refer Sec.\ref{sec:Deployment}). 
\item We collect a large number of traces from the app and systematically analyze them to detect TD of various services on the Internet (refer Sec. \ref{sec:Deployment}). 
\end{itemize}

\section{Related work}
\label{sec:rel_work}
Several methods have been proposed in the literature to detect Network neutrality violations as documented in recent surveys \cite{nn_survey_2018, nn_survey_2020}. The literature is rich with various forms of discrimination and its detection, like discrimination of content providers \cite{nnvd_cpdiscri_pls}, end-users \cite{nnvd_userdiscri_plr}, specific services like BitTorrent \cite{bttest}. Our focus in this work is detecting discrimination of a broad set of services that fall in the category of streaming services (both audio and video).  

Previous works have mainly used two methods to detect traffic differentiation using end-user side measurements, namely passive and active probing. \textit{NANO} \cite{nano} is a passive measurement tool. NANO compares the performance of a service on multiple ISPs and sees if its performance is degraded on an ISP to identify discrimination. A big challenge in NANO is identifying various confounding factors and normalizing their effect to make the comparison meaningful. Moreover, NANO runs in a centralized system, and the end-user cannot use it to detect traffic discrimination on their network.

\begin{figure*}[htbp]
	\centerline{\includegraphics[scale=0.13]{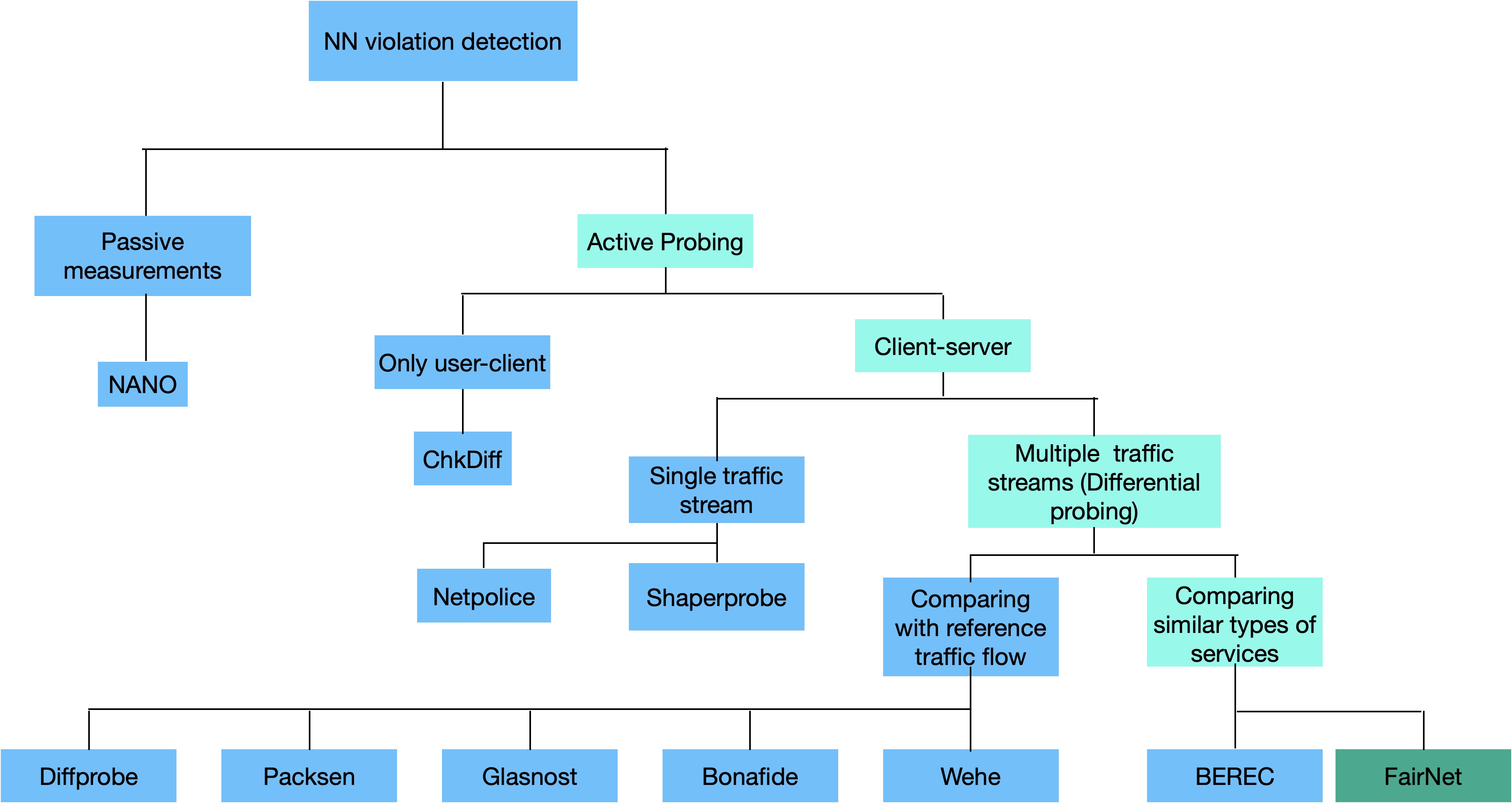}}
	\caption{Taxonomy of net-neutrality violation detection tools}
	\label{fig:nn_rel_works}
\end{figure*}  
Active-probing uses specially crafted traffic to remove the effects of many confounding factors such as network TMPs. Methods based on active probes sometimes generate a response from network middle-boxes, e.g., ICMP packets with specific values for the TTL field, to identify the source of discrimination. \textit{ChkDiff} \cite{chkdiff} is based on such active probes. However, such methods lack reliability as responses to active probes from the network middle-boxes are implementation-dependent. The dedicated client-server system used in our method removes this dependency. Sometimes, middle-boxes can identify active probes and may treat them differently. We ensure the correct classification of our probing traffic instead of being detected as probing traffic.

\textit{Netpolice} \cite{netpolice} and \textit{Shaperprobe} \cite{shaperprobe} use dedicated client-server for exchanging traffic stream. \textit{Netpolice} uses active probes as in \textit{Checkdiff}, whereas \textit{Shaperprobe} uses level-shift in received data rate crossing certain threshold as an indicator of the traffic discrimination/shaping. The effect of confounding factors such as congestion is mitigated in \textit{Shaperprobe} to a large extent by applying stochastic conditioning. Our method uses the service-type specific data delivery and the parallel data download mechanisms to mitigate such effects effectively. 

Other methods of detection of traffic discrimination are based on network tomography \cite{nn_infer} and measuring packet scheduler level parameters such as available tokens to detect traffic policing \cite{analysing_trpolicing}. Detecting only traffic policing in the context of net-neutrality violation detection is insufficient as the traffic flows, in practice, undergo various performance degrading effects such as congestion and traffic shaping. 

Differential probing is another method for detecting discrimination that compares two or more traffic streams with one of the streams specially crafted to act as a reference and indicate ISP's non-neutral behavior. \textit{DiffProbe} is one of the first tools based on this mechanism. It compares an application flow's delay and packet loss performance (A-flow) with a specially crafted probe flow (P-flow). The method's dependency on saturating the user's link is a limitation as it is rarely possible to achieve this. The other similar tool \textit{Packsen} \cite{packsen} performs the stochastic comparison of the inter-arrival times of packets and measured bandwidth of a 'baseline flow' and the 'measured flow' for detecting traffic shaping. The baseline flow is assumed to be non-shaped. However, the method does not consider the effect of traffic classification on packet treatment; instead, it relies on identical data transmission for receiving identical performance for both flows at the receiver. \textit{Glasnost} \cite{glasnost} compares performances of two flows of an application like BitTorrent that runs one after another. The second flow is the same as the first flow except that the protocol header parameter and application payload content is bit-reversed and act as a reference flow. This technique will not work for HTTPS traffic as the entire flow gets encrypted, and ISPs cannot classify based on payload. 

The most recent tools \textit{Bonafide} \cite{bonafide} and \textit{Wehe} \cite{wehe} are other examples of differential probing that uses original application data in payload and compares its performance against specially generated reference traffic. Note that both traffic streams maintain the timing relationship of the replayed HTTPS content. \textit{Bonafide} uses traffic with random data as reference, and \textit{Wehe} uses the VPN or bit-flipped version of the original service's data to generate traffic that acts as a reference which evades the traffic management performed by network middle-boxes. 

Experimental validations of limitation of the Wehe tool are done in \cite{nnvd_chls}. Fig.~\ref{fig:nn_rel_works} shows the taxonomy of existing net-neutrality violation detection tools contrasting the features of FairNet with others. Specifically, FairNet is based on active probing and uses the dedicated client and server design to normalize congestion, as considered by earlier works.

\subsection{Shortcomings of existing tools:}
Existing tools suffer from many shortcomings that make them less practical for detecting TD on HTTPS traffic. In the following, we discuss two of them.

\noindent
{\bf Mis-classification of services:} Almost all modern streaming services uses the HTTPS protocol to encrypt the payload for security and privacy reasons. However, previous tools do not make necessary changes in probing/replay traffic to be classified as original services, like establishing a secure channel using appropriate parameters for TLS handshake before exchanging application data for measurements. 

Due to the encrypted traffic and improper TLS handshake, middle-boxes cannot get appropriate signatures from either payload, application header, or initial TLS handshake, which may create ambiguity for the middle-boxes to classify the replayed traffic appropriately. Wehe is the latest tool for TD detection. It replays the original service's traffic while maintaining application-level timing relationships. Using a commercial traffic shaper, We demonstrate that Wehe's traffic is not classified as original service as required by the tool for TD detection. 

\noindent
{\bf Unreliable reference traffic:} Previous works use two types of references. Active measurements use a reference flow specifically crafted to evade discrimination. A weighted average of QoS received by several flows other than the test service is taken as the reference in passive measurement. In the first type, replay traffic is generated by modifying specific attributes of the original traffic, such as port numbers, randomizing payload, or access over VPN. Changing such attributes often changes the service type of replay traffic to be different from the original service, e.g., any change of port $443$ to any other non-standard port makes the HTTPS browsing traffic classified as P2P traffic. Comparison of performance of target traffic with such reference traffic will not lead to correct conclusions. Passive methods obtain reference performance by taking a weighted average of varied types of services. However, this is problematic as the network may treat different services with varied QoS on the Internet. Thus existing tools suffer from the issue of having reliable reference performance. 

FairNet measurement framework does not suffer either of these limitations as we exchange the HTTPS content on port $443$ and compares services of similar types. Also, any quality degradation on the Internet is not NN violations.  The study in \cite{nn_regu_violation_detect} shows that $39\%$ to $48\%$ of the detected traffic discrimination are not Net Neutrality violations according to regulatory instructions. Thus, there is a need for a tool that clearly detects  NN violations without ambiguity and conforms accurately to existing regulations, such as that recommended by BEREC\footnote{Body of European Regulators for Electronic Communications}. Our work aims to fill this gap.

The organization of paper is as follows. 
Sec.~\ref{sec:e2e} covers the implementation of FairNet measurement framework. Sec.~\ref{sec:results} explains the validity of its different components. Sec.~\ref{sec:Advantages} lists the operational limitations and advantages. The analysis of measured data generated by deployed FairNet framework is presented in Sec.~\ref{sec:Deployment}. The Sec.~\ref{sec:conclusion} concludes the paper.     

	
\section{FairNet measurement system}
\label{sec:e2e}
FairNet measurement framework consists of a user-client interacting with the ``common server" (CS). In the rest of this section, we describe the operations of each of its components.
\begin{figure}[htbp]
	\centerline{\includegraphics[scale=.15]{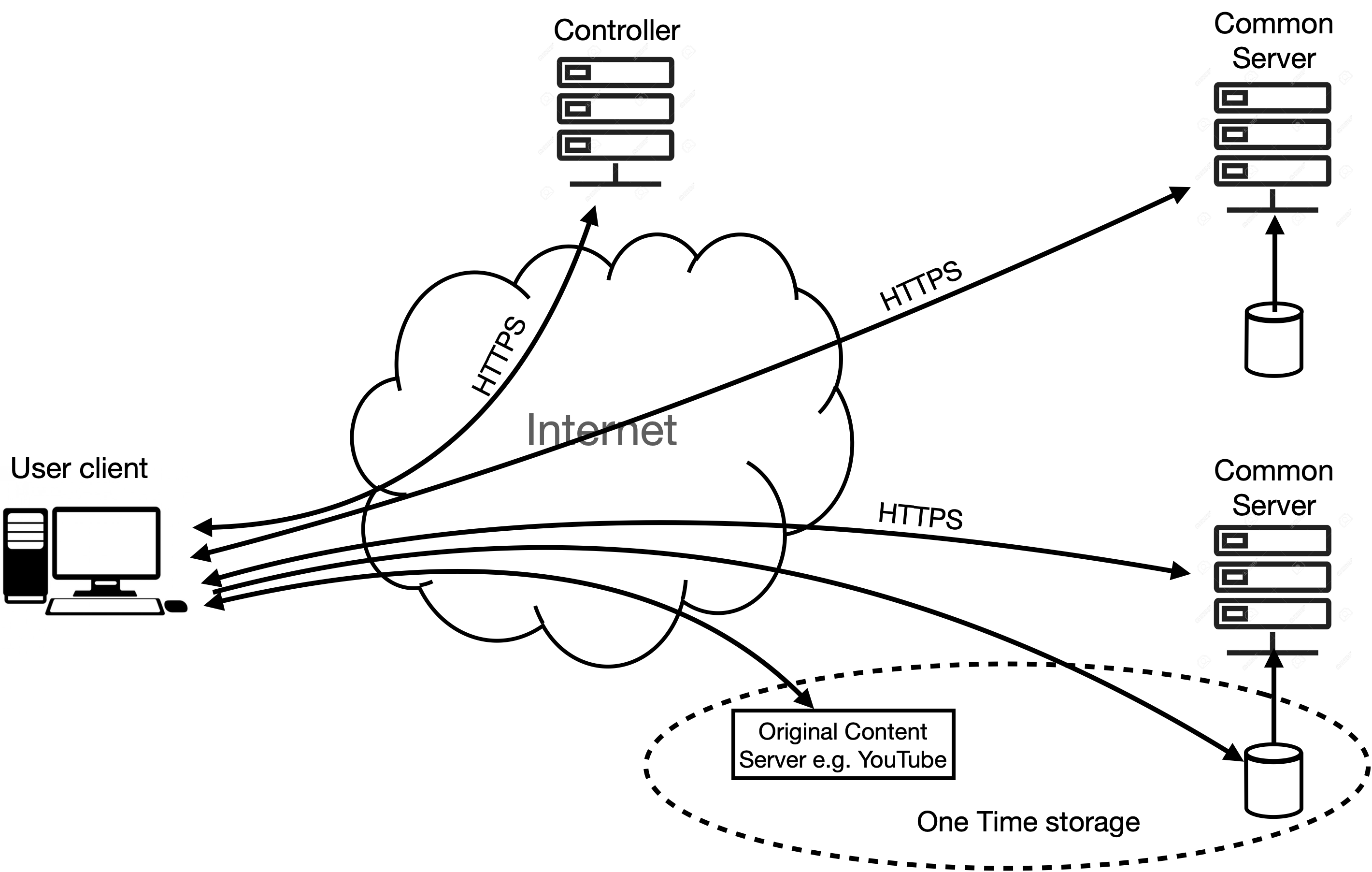}}
	\caption{End-to-end performance measurement setup}
	\label{fig:e2e}
\end{figure}

\subsection{Common Server (CS)}
\label{sec:comm_server}
The CS hosts the 20MB of application data for each service, extracted from the original service traffic logs, for sending to the user-client whenever requested. The original service accessed using a web browser generates the required traffic logs for individual services. The CS emulates the transmission strategy of the original service while transmission, e.g., the behavior of commercial streaming servers for transferring data using customized Dynamic Adaptive Streaming over HTTPS (DASH). 

Each service is treated independently at the server-side, thanks to a per-service different connection mechanism design. This way, the server design is scalable to support any number of users without adding book-keeping overhead. The CS stores the application data of each supported service in small chunks called segments, and each segment needs a separate request from the user-client. The content of each segment is further divided into smaller-sized bursts for transmission to make efficient use of network queues and their scheduling mechanisms. The server performs the burst size adaptation at the start of the transmission of each segment. It modifies burst size in response to observed throughput for a given service under test.  

 \noindent
\textbf{Customised DASH:}
Our method uses a generic rate adaptation mechanism to adapt all the services' data rates with the same base rate. The rate adaptation avoids the data rate overshoot available bandwidth in the system. It also ensures that the end-user performance is similar for all active services for a given user under a similar congestion environment. Thus avoids the inconsistencies in the end-user performance measurements. Any such inconsistency, if present, indicates the routing of services through different paths, probably some of them being more congested. We also verified that the common transmission strategy does not impact services' correct identification and classification using a commercial traffic shaper. Details of customized DASH used are  given in Appendix  \ref{appdx:dash} and its validation on classification of traffic are given in Section \ref{sec:css_os_tr_similarity}.

\subsection{User Client}
\label{user_client}
User client provides an interface to connect to the CS. Our method works on performance comparison of similar services. The client provides an option to select from a list of supported services. In the client, one selects a service to be tested for discrimination (test service). The client then randomly selects another service (reference service) that having similar QoS requirements. For example, if the user client needs to test if there is degradation in YouTube's quality, the method selects services like Netflix or Prime Video as reference services. 

The client sets up HTTPS connections for each selected service with CS simultaneously. In the ``clientHello" message of the initial Transport Layer Security (TLS) negotiations sent for each service, the user-client sets the appropriate value of the Server Name Indication (SNI). We obtain SNI for each service from the network logs of data download from the original servers. As SNI information goes in plain text, it is easier to read it from the intercepts. 

The user client downloads the content of selected services from the CS. The download process stops either if $20$MB content is received or the download continues for $3$ minutes. It records application-level data received for each service along with its timing reference during download. The algorithm described in Section~\ref{sec:trdiffdetect} uses this data to detect discrimination. The outcome of TD detection, along with a comparative throughput plot, is presented to the user for further investigation.

\noindent
{\bf Mobile Apps:} The user-client used in our measurement framework is available as an application for both Android and iOS devices. Both Apps do not require any permission from the user for their execution. We only ask consent from the user for gathering measurement reports. The minimum supported Android OS version is Android "Nougat" (Version-7). The use of the SNI field in the initial SSL client hello message enforces this restriction. iOS App requires version 14.4 and above.
The FairNet App supports popular video and audio streaming services like Netflix, Prime Video, YouTube, Hotstar, MXPlayer, Hungama, Zee5, Voot, EROS Now, SONY Liv,  Spotify, Saavn, Wynk, PrimeMusic, and Gaana.

\begin{figure*}[htbp]
    \centerline{\includegraphics[width=4in]{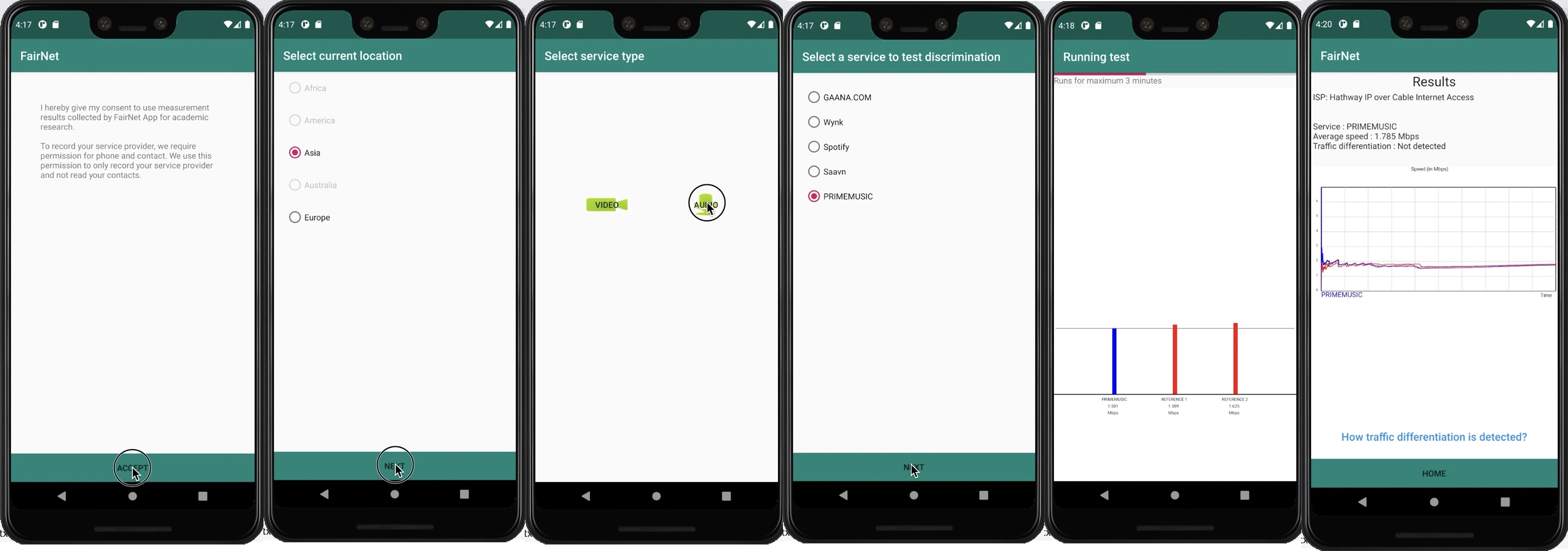}}
    \caption{FairNet App on Android}
    \label{fig:fnapp}
\end{figure*}

The Apps are interactive, asking user input and presenting the progress of the test as well as the conclusion, as shown in Fig. \ref{fig:fnapp}. During the test, the dynamic bar chart shows the visual representation of the cumulative throughput of each service selected by the user. On completion of the test, the App shows the verdict indicating the presence or absence of TD.

\noindent
{\bf Web-based client:} The Java-based web client is another option for providing end-user FairNet clients. The advantage of using a web-based client is its accessibility from a web browser, making it suitable for any device that supports java based web browsers. Java-applet and javascript-based HTML web-URL are two options to develop such a client. 

The java-applet execution needs NPAPI technology. The Chrome browser no longer supports it. Thus it requires specific browser settings and the installation of particular extensions to run the Java applet in Chrome. Other browsers also require particular settings to run Java applets. Moreover, the requirement of setting the original server's SNI in TLS handshake while completing the handshake using a self-signed certificate is not possible with publicly available browsers such as Chrome. Hence instead of a web-based client, we use the mobile app for deployments in the wild.

\subsection{Traffic Discrimination (TD) detection}\label{sec:trdiffdetect}
\label{sec:tddetectalgo}
TD can be induced in the network either by a direct means, e.g., throughput is restricted/throttled, or subtly, e.g., service connections are reset or gets regular rapid throughput variations. To cater to the broad spectrum of induced TD in the presence of network effects, we develop two types of specialized detection algorithms.
The two algorithms are,
\begin{itemize}
	\item Throughput Consistency Detection (TCD)
	\item Connectivity Status Detection (CSD)
\end{itemize}

The user-client performs TD detection after it finishes $20$MB sized file download for all considered services or after the lapse of $3$ minutes, whichever happens earlier. Download size of $20$MB is selected empirically to allow measurements to reach stable throughput values (refer Fig.~\ref{fig:comp_diff_types}). 
Our algorithms use average throughput calculated over predefined slots of fixed duration. The entire duration of the test is represented by the number of total non-overlapping time slots $N_T$. The time slot is a configurable variable that is fixed experimentally by observing its effect on detection capability using a validation setup explained in Sec. \ref{sec:tdalgocal}.  
\subsubsection{Throughput Consistency Detection}
This algorithm detects TD based on the observation that the throughput performance of discriminated (through malicious TMPs) services remains low for most of the streaming session. It is a throttled/low throughput duration, and the algorithm precisely detects this duration and compares it with the other services selected for comparison. The throttled duration comparison strategy enables the algorithm to handle normal as well as delayed throttling\footnote{The delayed throttling occurs when ISP does not apply throttling immediately after starting data download but applies it after some time during the download.}.  Moreover, the throttled duration comparison helps detect any presence of throttling, randomly distributed over the entire duration.  

\begin{figure}[htbp]
	\centerline{\includegraphics[scale=.13]{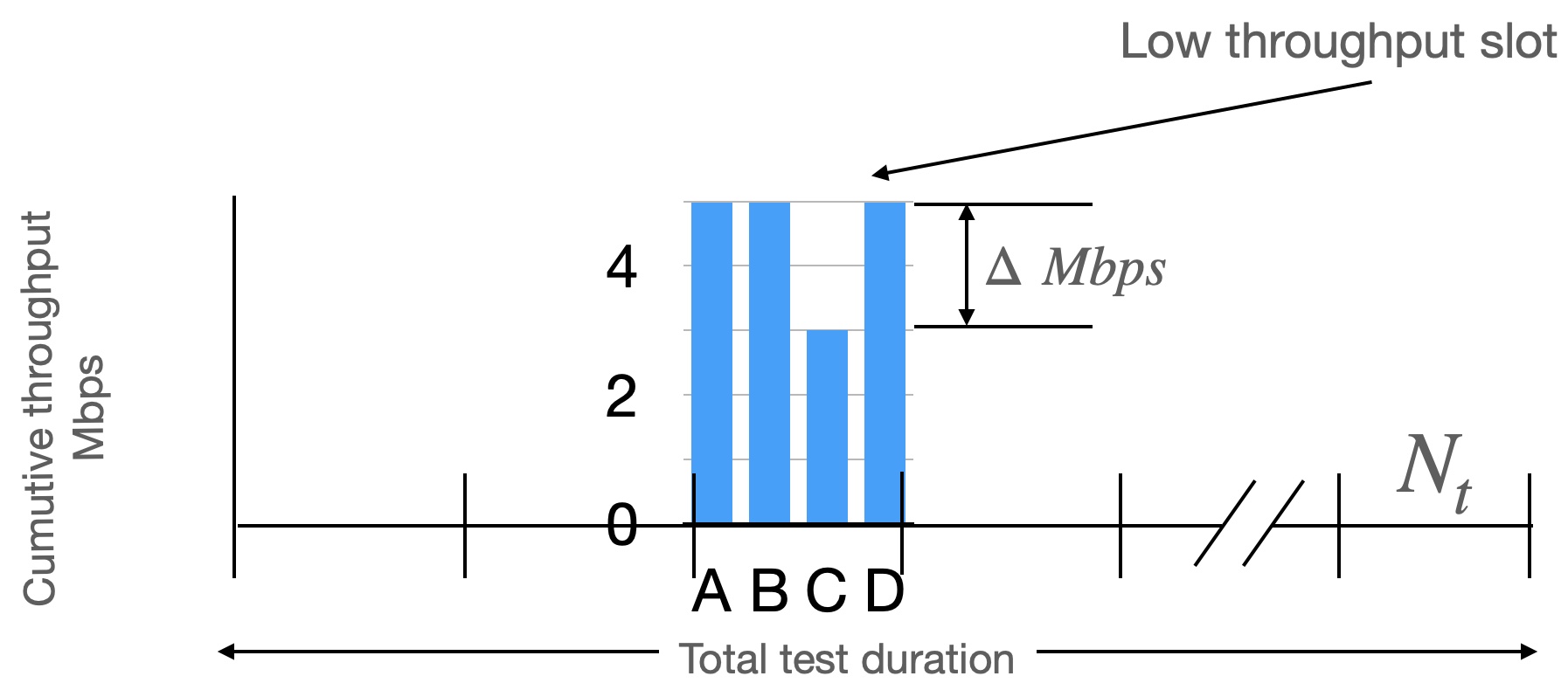}}
	\caption{Low throughput slot marking}
	\label{fig:lth_slot}
\end{figure}

\begin{figure}[htbp]
	\centerline{\includegraphics[height=4cm, width=8cm]{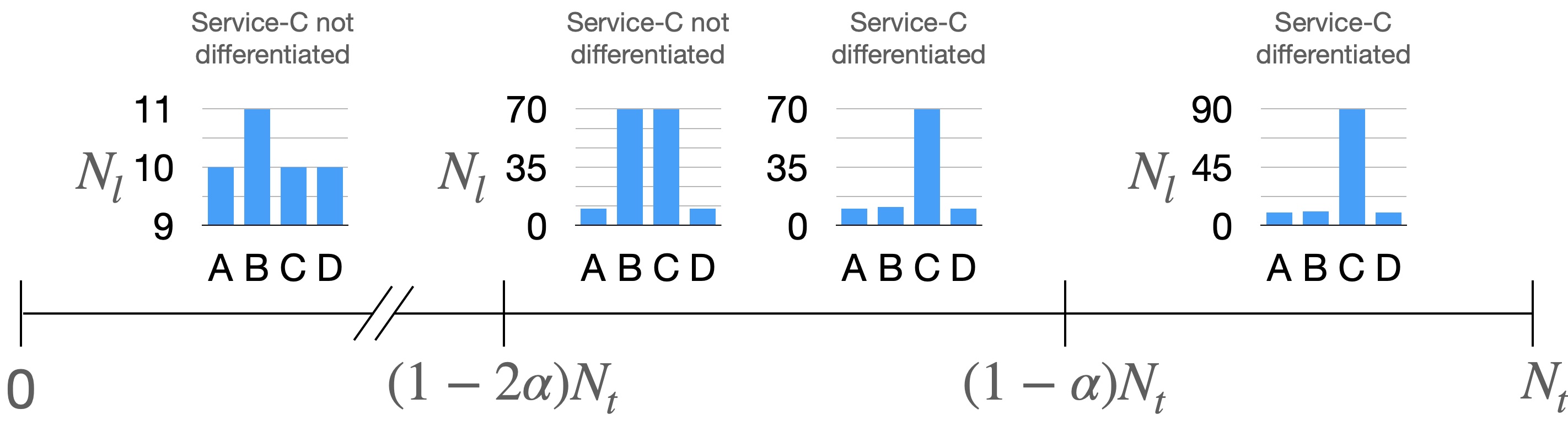}}
	\caption{Working of TCD algorithm}
	\label{fig:tcd}
\end{figure}

Fig.~\ref{fig:lth_slot} and \ref{fig:tcd} shows the working of the TCD algorithm that illustrates the TD detection performed for a service named `C'. The Algo.~\ref{algo:TCD} gives the pseudo-code of the algorithm. A slot is tagged as a `low throughput' for a service if its average throughput is at least $\Delta$ Mbps below the highest throughput recorded in that slot. The algorithm counts the number of low throughput slots for each service. Let $N_L$ denotes the number of low throughput slots for the service under test in Algo~\ref{TrDiffDet-th}. The algorithm checks if $N_L$ is more than $(1-a)N_T$ slots. If true, then the test service is declared as discriminated for throughput. If this is not the case, a second softer threshold on $N_L$ is checked to avoid missing detection (false-negative) due to the hardbound. If $N_L$ is in the range $(1-2a)N_T \mbox{ to } (1-a)N_T$, then the algorithm checks for how many services' $N_L$ also falls in the range $(1-2a)N_T  \mbox{ to } (1-a)N_T$. This count is denoted as $N_s$ in Algo.~\ref{algo:TCD}. If $N_s$ is more than one, then possibly the service under test is not the only one experiencing lower throughput, and no throughput discrimination is reported. Otherwise, service under test is declared discriminated. The rationale behind using the $N_s$ condition during the softer threshold is to ensure that TD detection is not falsely detected if multiple services, having similar base rate and its DASH (refer Sec.~\ref{sec:comm_server}), are experiencing throughput performance within softer thresholds. It is a case of lower throughput for multiple services due to wrong momentary re-routing through relatively congested routes. 

As mentioned above, the algorithm is a differential detector and needs to compare the performance of the service under test with other services. If ISP discriminates against all services selected by the client, then our method will not declare any discrimination.  
\begin{algorithm}[!ht]
	\caption{Throughput Consistency Detection (TCD)}\label{TrDiffDet-th}
	\label{algo:TCD}
	\begin{algorithmic}[1]
		\State {\bf Input:} $N_T,N_L, N_s, a$ 
		\If{$N_L \ge (1-a)N_T$}
		\State Throughput discrimination TRUE
		\ElsIf {$(1-a)N_T< N_L \le (1-2a)N_T$}
		\If{$N_s > 1$}
		\State  Throughput discrimination FALSE
		\Else
		\State  Throughput discrimination TRUE
		\EndIf
		\Else
		\State  Throughput discrimination FALSE
		\EndIf
	\end{algorithmic}
\end{algorithm}       
The Sec.~\ref{sec:tdalgocal} describes the calibration method to set thresholds (slot and throughput) and slot time used in the TCD algorithm.

\subsubsection{Connectivity Status Detection}
A malicious ISP can disturb the client-server communication by breaking the end-to-end connection abruptly mid-way, forcing user-client to re-establish the connection. Such a phenomenon could degrade the user experience even though such disruption results in discarding not much data. The ISP-induced RESET is one common reason for such breaks. Our algorithm detects such scenarios based on connectivity status. The Algo. \ref{TrDiffDet-CB} gives the pseudo-code of the CSD algorithm. It detects TD by comparing the number of connectivity breaks against a fixed threshold $b$. It is based on the fact that the average segment size used by streaming services is $2$MB. This results in $10$ segments for $20$MB data. The threshold of $5$ indicates connectivity breaks for $50\%$ of segments. The threshold of $5$ emphasizes continuous connectivity breaks, possibly due to  ISP's malicious behavior.     

\begin{algorithm}[!ht]
	\caption{Connectivity Status Detection (CSD)}\label{TrDiffDet-CB}
	\begin{algorithmic}[1]
		\State {\bf Input} $b, N_cb$
		\If {$N_cb <  b$}
		\State Connectivity discrimination FALSE
		\Else 
		\State Connectivity discrimination TRUE
		\EndIf
	\end{algorithmic}
\end{algorithm}       

We do not proceed with TD detection if none of the services could finish the file download during maximum download duration, i.e., $3$ minutes. It represents the ``bad network" conditions for making comparable measurements. Otherwise, we perform both TCD and CSD. Table~\ref{tab:tdde} shows how we infer TD using the outcome of both TCD and CSD as given in Table~\ref{tab:tdde}. 

Throughput performance of traffic stream reflects how its packets are treated over the Internet. So FairNet uses directly comparable throughput performance as a primary indicator of TD. However, if the TCD algorithm does not detect TD, the outcome of the CSD algorithm is taken into consideration. The CSD algorithm detects TD due to a large number of connection breaks. Such a large number of breaks may not result in throughput reduction. However, it can force repeated re-fetching of application data, prohibiting completing complete $20$ MB data. Thus TD is declared detected based on CSD algorithm outcome in this case. However, TD detection continues to follow the outcome of TCD if data download completes indicating less number of data re-fetches. TD is declared to be not detected if both algorithms do not detect it.

\begin{table}[!t]
    \renewcommand{\arraystretch}{1.3}
    \caption{TD detection preference}
    \label{tab:tdde}
    \centering
    \small
	\begin{tabular}{|c|c|c|c|}
		\hline
		\begin{tabular}{c}
			TCD \\
			outcome
		\end{tabular}  & 
		\begin{tabular}{c}
			CSD \\
			outcome
		\end{tabular} & 
		\begin{tabular}{c}
			Test service \\
			Download
		\end{tabular}  & 	\begin{tabular}{c}
			TD \\
			Detected?
		\end{tabular}   \\
		\hline
		TRUE & TRUE & NA & YES\\ 	\hline
		TRUE & FALSE & NA & YES \\ 	\hline
		FALSE & TRUE & 
		\begin{tabular}{c}
			$100\%$ \\ 	\hline
			$< 100\%$
		\end{tabular} & 
		\begin{tabular}{c}
			NO \\ 	\hline
			YES
		\end{tabular}\\	\hline
		FALSE & FALSE & NA & NO\\
		\hline
	\end{tabular}
\end{table}

\subsection{TD detection algorithm calibration}
\label{sec:tdalgocal}
\begin{table}[!t]
    \renewcommand{\arraystretch}{1.3}
    \caption{TCD algorithm Calibration}
    \label{tab:tcd_calib}
    \centering
	\begin{tabular}{|c|c|c|c|}
		\hline
		\begin{tabular}{c}
			Throughput \\
			threshold
		\end{tabular}  & 
		\begin{tabular}{c}
			Slot \\
			threshold
		\end{tabular} & 
		\begin{tabular}{c}
			Slot \\
			time
		\end{tabular}  & 	\begin{tabular}{c}
			\%TD  \\
			error
		\end{tabular}   \\
		\hline
		1.75 & 0.2 & 1 & 4.14 \\ 	  \hline
		1.75 & 0.3 & 1 & 1.18 \\ 	  \hline
		1.75 & 0.2 & 1.75 & 2.95 \\ \hline
		\rowcolor{LightCyan}
		1.75 & 0.3 & 1.75 & 0.29 \\ \hline
		1.5 & 0.2 & 1 & 2.21\\ 	  \hline
		1.5 & 0.3 & 1 & 0.73 \\ 	  \hline
		1.5 & 0.2 & 1.75 & 1.03 \\ \hline
		1.5 & 0.3 & 1.75 & 1.03 \\
		\hline
	\end{tabular}
\end{table}

FairNet's TD detection algorithm uses two threshold values to compare the performances of different services, viz throughput and slot threshold. We used a test-bed in a lab environment to calibrate the detection algorithm. The test-bed consists of a FairNet client and server on a single machine to reduce any network noise. We vary the throughput of specific services using throughput attenuating logic attached to the server. This environment ensures noise-free simulation of TD through the predefined adequate difference in the throughput performances of test services without any ambiguity.

\noindent
\textbf{Calibration method:} We varied the throughput threshold, slot threshold, and slot time within an acceptable range in the detection algorithm to analyze its performance on thousands of network logs. Each network log corresponds to a combination of $8$ video streaming and $4$ audio streaming services. We denote the total number of scenarios as $N_t$. Almost $50$\% of network logs contains simulated TD. We denote the number of network logs with TD as $N_{ip}$ and the number of TDs detected by the algorithm as $N_{cal}$. Then we defined the TD classification error as $(N_{ip}-N_{cal})/N_{ip}$. The classification error is calculated for each combination of the both thresholds using different slot times.  \\
\noindent
\textbf{Calibration criteria and outcome:} For all tried combinations, we looked for the combination with the lowest TD classification error. Our experimental results for some combination of parameters are summarised in Table~\ref{tab:tcd_calib}. We tried over a range of values of the parameters varying throughput in the range $1.5$-$1.75$ Mbps, slots in the range $0.2$-$0.3$, and slot time in the range $1$-$1.75$ seconds. As seen, we found that the combination with throughput threshold of $1.75$ Mbps, slot threshold of $0.3$, and slot times of $1.75$ seconds results in the lowest TD classification. We thus use these values in our TD detection algorithm.
	
\section{Validity of  Framework}
\label{sec:results}
The success of our measurement framework depends on the following two points: 1) the services selected for comparison experience similar network effect so that their performance is directly comparable 2) the replay traffics are correctly classified and treated as the original service's traffic, e.g., replayed YouTube traffic from the CS should be classified as YouTube service traffic in the network. In this section, we validate both of these points using commercial traffic shaper. We also validate that our TD detection mechanism is robust against various network fluctuations.

\subsection{Comparison of services with similar QoS requirements}
\label{subsec:perf_compare}
FairNet performs TD detection using the comparison of multiple streams of similar types. A fair comparison of the replayed traffic is possible only after normalizing the effect of any TMPs. The expectation is that ISPs treat services requiring similar QoS with similar TMPs. Reports from Body of European Regulator for Electronic Communication (BEREC) \cite{berec_9277, berec_7296, berec_7243} also recommends comparison of streams of similar QoS requirements as a means to detect if one service is discriminated against the other. Specifically, \cite{berec_1101}[Sec 5.3.1] states that "If the application under consideration receives a significantly lower throughput than similar applications, this may signify a case of application-specific throttling." Our method is in line with this recommendation.

\begin{figure}[htbp]
	\centerline{\includegraphics[scale=.275]{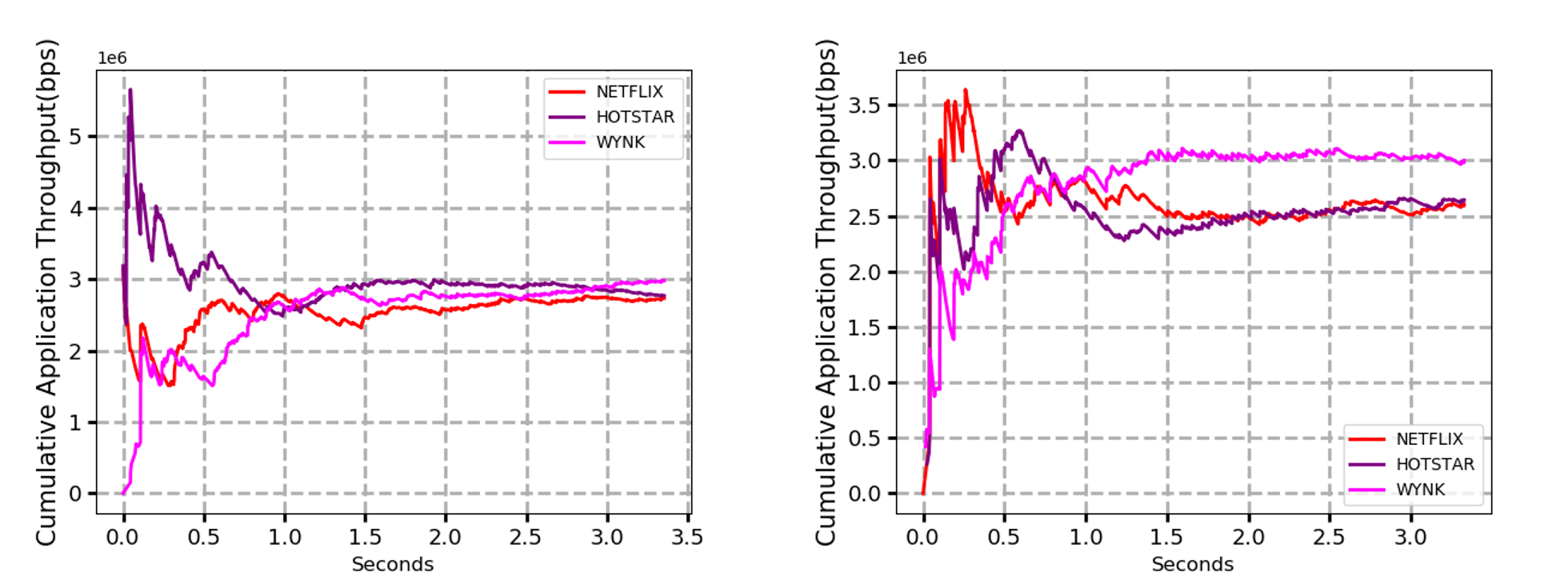}}
	\caption{Performance comparison of services with different QoS requirements e.g. video and audio streaming}
	\label{fig:comp_diff_types}
\end{figure}

To validate that services requiring similar QoS are subjected to the same TMPs in the network, we compared the throughput of Netflix, Hotstar, and Wynk services. Netflix and Hotstar are video streaming services having similar QoS requirements, while Wynk is an audio streaming service with different QoS requirements. Fig.~\ref{fig:comp_diff_types} shows the throughput of these services measured on the Internet at two different times. While the left-side plot shows similar performances for all services, the right-side plot shows variations in it. Similar performances can be attributed to sufficient bandwidth availability. However, though the quality of services varies with time, variations for video services are identical in each plot and are different from audio streaming service Wynk. The primary reason being they are treated the same way in each of the measurements as they have similar QoS requirements, whereas the audio service with different QoS requirements is treated differently. As we compare a test service with other services having similar QoS requirements, all the services under comparison in our measurement setup are likely to undergo similar network effects, thus normalizing its effect. 
\begin{figure*}[!t] 
	\centering
		\subfigure[Setup for Validation of Measurement Framework using traffic shaper\label{fig:tsv}]{
		\includegraphics[width=0.45\linewidth]{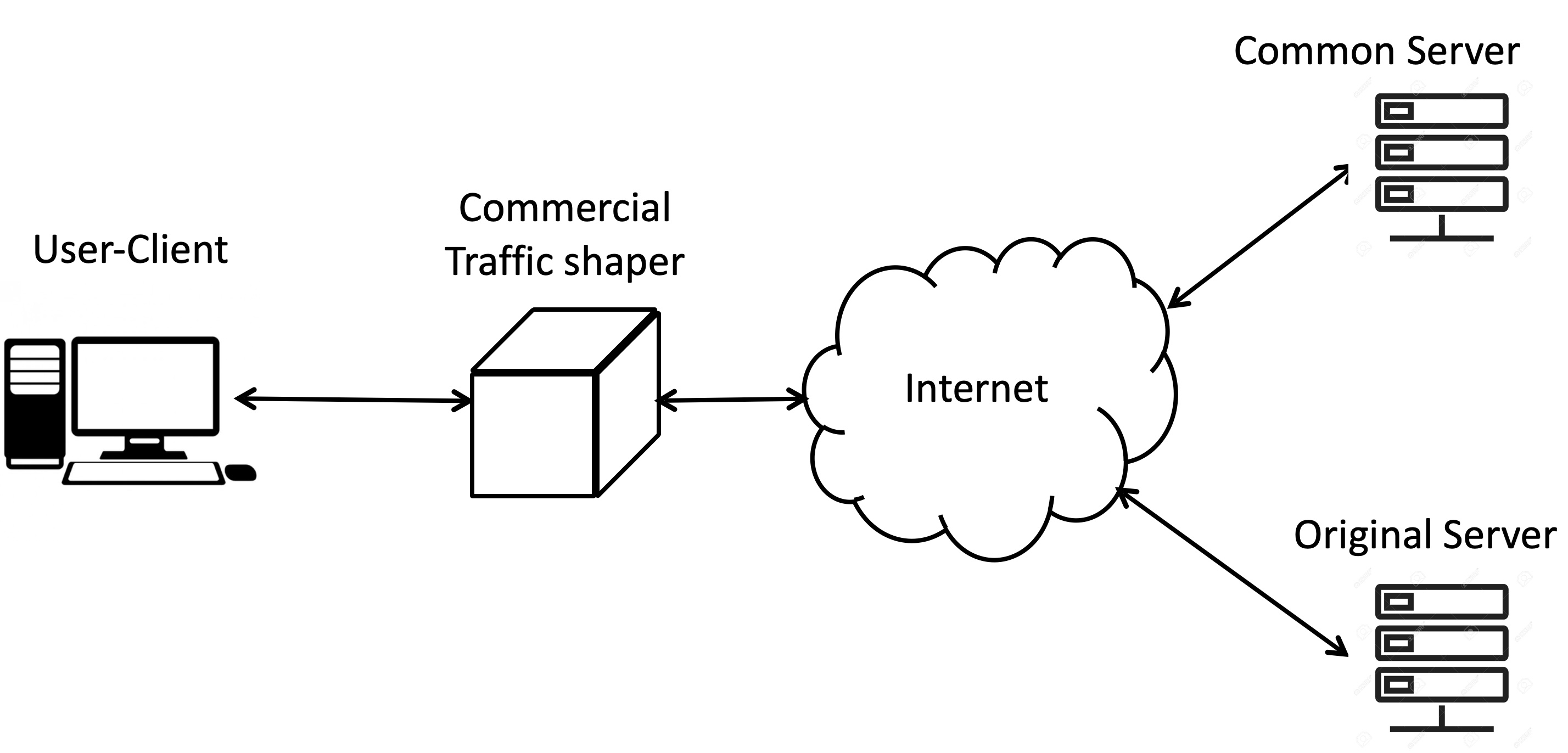}}
	\subfigure[Using internet browser\label{fig:yt_intbrow}]{
		\includegraphics[width=0.45\linewidth]{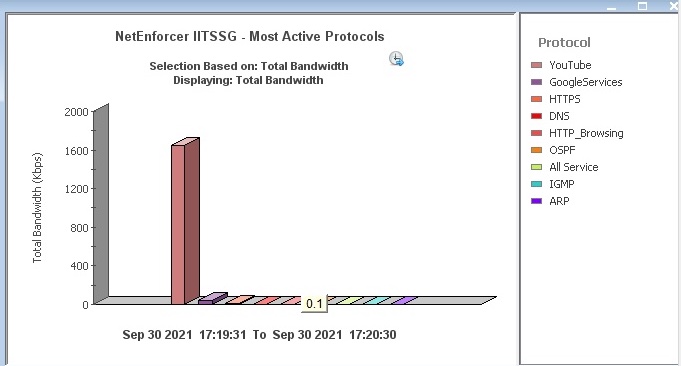}}
	\subfigure[Using user-client script (without SNI )\label{fig:yt_withoutsni}]{
		\includegraphics[width=0.45\linewidth]{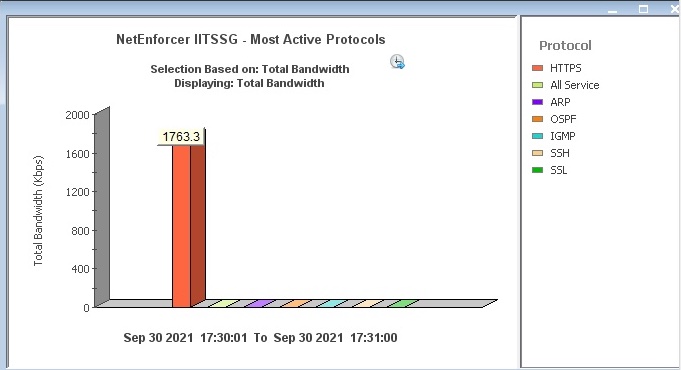}}
	\subfigure[Using user-client script (with SNI )\label{fig:yt_withsni}]{
		\includegraphics[width=0.45\linewidth]{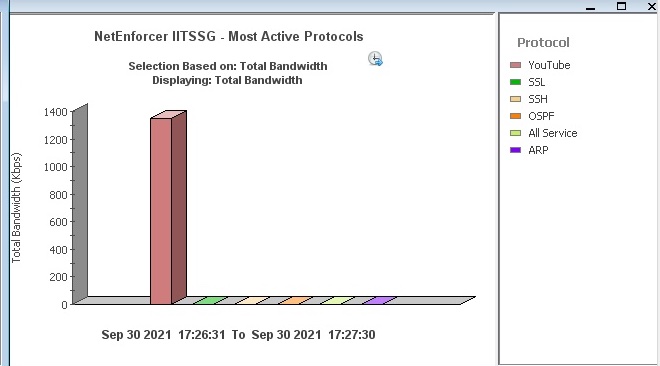}}
		\caption{Traffic classification results from the traffic shaper when YouTube is accessed directly from web-browser and from user-client script.}
	\label{fig:yt_trcl}
\end{figure*}

\subsection{Similarity of Traffic from CS and Original Server}
\label{sec:css_os_tr_similarity}
We next validate if the replayed traffic from the CS is classified as intended traffic by the network middle-box. As most services use the HTTPS protocol, traffic classification is challenging due to encrypted application payload. Classical traffic identification mechanisms using DNS information or TCP header information are unreliable due to the advent of DNS over TLS technology and dynamic address/port allocation techniques. To overcome this, ISPs employ advanced techniques like Deep Packet Inspection (DPI), and Deep Flow Inspection (DFI) \cite{tr_cl} for classification of HTTPS traffic. DFI-based techniques using traffic characteristics are often unreliable as traffic characterizations changes as services adapt to different technologies and network conditions. However, DPI based technique using Server Name Indication (SNI) \cite{tls_rfc7858} parameter is more reliable as SNI is exchanged in pain-text on port $443$ in the initial TLS handshake. Since SNI is easy to intercept, it is one of the preliminary checks for traffic classification and used along with other techniques for traffic classification \cite{cisco-sdavc}. Exploiting SNI information also offers higher accuracy \cite{trcl_survey} in traffic classification.

To study the role of SNI in influencing the classification result of middle-boxes, we use a setup as shown in Fig.\ref{fig:tsv}. In this setup, a user-client can access the service's content from the CS using scripts or access it directly from the original server using a web browser through a commercial traffic shaper in both cases. For example, the user-client can directly access YouTube content from YouTube.com using a web browser or connect to the CS to download replay traffic of YouTube using a script. During data transfer, the traffic classification activities by the commercial traffic shaper are recorded from its User Interface (UI) (see Figs.~\ref{fig:yt_intbrow},\ref{fig:yt_withoutsni},\ref{fig:yt_withsni}). We ensured that there was no other major Internet activity during experiments.

We wrote a script to access replay content from the CS. The script performs the same job as web-browser but gives us control over setting the SNI parameter. We performed experiments with the following three scenarios:
\begin{enumerate}
    \item YouTube.com is accessed using a web browser from the user client
    \item User client accessed YouTube replay traffic from the CS using the script. The script does not send SNI information in the TLS handshake\footnote{YouTube accepts TLS handshake without SNI parameter. However, other services like Netflix did not allow this}.
   \item User client accessed YouTube replay traffic from CS with proper SNI parameters in the TLS handshake. The SNI parameters are extracted from the original YouTube server's traffic logs.
\end{enumerate}
The classification outcomes of the shaper for the three scenarios are shown in Figs.~\ref{fig:yt_intbrow},\ref{fig:yt_withoutsni},\ref{fig:yt_withsni}. As expected, the shaper correctly classified the traffic in the first scenario as seen from Figs.~\ref{fig:yt_intbrow}. In the second scenario,  it classified the replay traffic as generic HTTPS traffic and not as YouTube traffic as seen from \ref{fig:yt_withoutsni}.
The classification outcome for the third scenario is shown in Fig.~\ref{fig:yt_withsni}. As seen, the commercial shaper correctly identifies the replay traffic as YouTube. The outcome of these experiments demonstrates that the network middle-boxes may not correctly classify the traffic only based on traffic characteristics imparted by the server; they look into the SNI as one of the primary parameters. Our experiments also confirm the observation in Cisco's report \cite{cisco-sdavc}.

\begin{table*}[ht]
\caption{Transport protocol and plaint-text server information}
\label{tab:tpservname}
\small
\centering
	\begin{tabular}{|c|c|c|c|c|c|}
		\hline
		Protocol & \begin{tabular}{c}Dominant \\Application \end{tabular}& \begin{tabular}{c} Secure channel \\ protocol \end{tabular} & Server Information & Support \\
		\hline
		\begin{tabular}{c} DCCP (Datagram Congestion \\ Control Protocol) \end{tabular}& media streaming & \begin{tabular} {c} DTLS \\ (Datagram-TLS) \end{tabular}& Not present & No \\
		\hline
		\begin{tabular}{c} TCP \\ (Transport Layer Protocol) \end{tabular} & \begin{tabular}{c}Almost every \\ secure service \end{tabular}& \begin{tabular} {c} TLS \\ \end{tabular}& \begin{tabular}{c}present (SNI in \\ "clienthello")\end{tabular}& Yes \\
		\hline
		\begin{tabular}{c} MP-TCP \\ (Multipath-TCP) \end{tabular} & Video streaming & TLS & present & \begin{tabular}{c} Yes  (single \\ network case)\end{tabular}\\
		\hline
		\begin{tabular}{c} UDP \\ (User datagram protocol) \end{tabular} & Real-time video & DTLS & Not present & No \\
		\hline
		\begin{tabular}{c} SCTP (Stream Control \\ Transport Protocol) \end{tabular} & Control channel & TLS & present & Yes\\
		\hline
		\begin{tabular}{c} MQTT \\ (MQ Telemetry Transport) \end{tabular} & IoT services & TLS & present & Yes \\
		\hline
		\begin{tabular}{c} QUIC (Quick UDP \\ Internet Connection) \end{tabular} & Video streaming & In-built TLS & present & Yes \\
		\hline
	\end{tabular}
\end{table*}

\subsection{Transport protocols and Server name information}
TD detection in FairNet relies on correctly classifying replay traffic from the CS by appropriately setting the SNI parameter in the secure channel establishment. The transport layer protocol dictates the protocol for secure channel establishment, e.g., TLS is used with TCP while DTLS is used for UDP transport. Thus the applicability of FairNet methodology for a given service is dependent on the transport layer protocol used by that service.  

The primary focus of FairNet is on multimedia streaming services primarily using TCP-TLS protocols. However, these services use various other transport protocols such as Multipath TCP (MPTCP) \cite{mptcp}, and QUIC \cite{quic}. Transport protocols other than QUIC send the TLS handshake data that reveals the server identity without any encryption. The FairNet framework works well on those protocols. QUIC obfuscates the initial Client-Hello message that includes SNI using an initial secret. However, the key can be retrieved from the received data using the standard defined operations and parameters that use SHA hash generation.  In effect, SNI information is available in plain-text ready form, and hence FairNet can be adapted to the QUIC protocol.

Table~\ref{tab:tpservname} lists the applicability of FairNet to major transport protocols. As server name information is readily available for TLS protocol in handshake messages, FairNet is readily applicable to services using TLS protocol for the secure channel establishment like TCP, MP-TCP, SCTP, MQTT, and QUIC. However, the absence of server name extension in DTLS makes the FairNet framework not applicable to transport protocols using it, e.g., UDP.
	
\section{Limitations and Advantages}
\label{sec:Advantages}
In this section, we discuss the advantages and operational limitations of the FairNet framework.

\subsection{Advantages}
FairNet overcomes the shortcoming of earlier tools and offers advantages that make it easy to use and scale better than the existing tools.  

\subsubsection {Seamless data collection:}
The user client collects data and performs measurements at the application layer. It simplifies the data collection in the mobile device as it does not require any special permission, e.g., rooting the device. Moreover, it does not require external system support for performance metric generation, e.g., VPN services required in Wehe.

\subsubsection{Scalable client management:}
The server manages each socket connection from a  client independently. The client keeps track of its download status and completes the processing of all required traffic flows. It makes the server free from associating sockets with user clients. Thus, available resources like RAM size, the number of parallel sockets the server can handle at the OS level, and the total available bandwidth can only restrict the client management in FairNet.

\subsubsection{No network traversing issues:}
The user client and CS communicate over standard socket connection using the TCP transport protocol. It is a standard configuration used by almost all Internet services. Thus, unlike Wehe, our measurement framework does not face difficulty connecting to any user client working under any network configuration, e.g., working behind a firewall or NAT device.  

\subsubsection{Streaming data in parallel:}
Since each service uses a different socket connection for data download, the user client can launch multiple simultaneous connections that stream data for multiple services simultaneously. The main advantage of such data transfer is ensuring similar network conditions to all services requested by a user client, thus facilitating a fair performance comparison. Note that previous works such as Wehe performs `back-to-back replay' of reference and application data pair, thus exposing both the streams to potentially different network conditions.

\subsubsection{Single sockets for each service:}
Our measurement framework allows the use of a single socket connection for each service. It removes the task of identifying and separating individual service data. 

\subsubsection{CDNs within ISP network:}
Many times content servers have an arrangement with ISPs for hosting their server inside ISP network \cite{cdn-isp-coll}. It improves the end-user quality of experience. Ultimately, such arrangements provide these services edge over other services. It is unfair to compare the performance of such services with other services in the context of TD. The CS removes this issue. It facilitates the fair comparison of performances of streaming services. 

\subsection{Operational Limitation}
One of the crucial aspects of FairNet is to use appropriate SNI parameters so that the replay traffic is correctly classified in the network. Many services change their SNI dynamically. The network middle-boxes periodically update their databases to respond to such changes. Thus it requires the FairNet to periodically update its SNI database to keep in sync with the original services' SNIs. 

Many times SNI values are region-specific. The traffic stream not using such region-specific values may be blocked/wrongly classified by ISPs that filter region-specific SNIs. It was observed in the experiment testing Netflix service from the CS located in Europe but using India-specific SNI.

An upgrade called Encrypted SNI (ESNI) \cite{esni} is being discussed and is in `experimental phase' to address the risk of domain eavesdropping from SNI. ESNI feature in TLS1.3 can make the FairNet solution ineffective if implemented. However, practical deployment of ESNI is a long-drawn process as research on its practical implementation is still ongoing.

\section{Deployment}
\label{sec:Deployment}
We deployed the measurement framework publicly using the FairNet App with a server running in Mumbai (Asia). We fine-tuned the server to the respective hardware resources available using experimental trials. As our method require gathering local SNI, we restricted our measurement to collecting network logs within India. As noted earlier (Sec. \ref{user_client}), FairNet is supported for implementation in mobile apps but not as a web application due to browser support issues. Collecting the traces from mobile app across multiple locations is a tedious task. It cannot be automated and has to be done manually. We requested several students residing in India to use the FairNet app to generate measurement traces and collect more than $1300$ traces over $12$ months.  

FairNet is developed keeping end-users in mind. It enables them to check if their traffic (over wireless network) is subject to discrimination. Hence not being supported through a web browser is not a concern. However, not being supported through browsers makes it difficult to run it on nodes on the Internet to automate the collection of measurement traces. In the following, we report our observations from the traces. 

Each log contains the unique user identity (consists of the timestamp to identify logs from the same user) collected from the user device, the ISP used for testing, and service performance data per service. The data includes the sequence of application-level received data size and its timestamp and the number of connection breaks during the test. 

These logs are analyzed offline to generate the comprehensive TD detection report. Even though the user-test generates the network logs for detecting the TD of a specific service chosen by the user, we analyzed the test logs for the TD of other services selected for comparisons. Our measurement framework detects TD characterized by lower throughput or frequent connection breaks for a given service using TCD and CSD algorithms and combing their results as given in Table \ref{tab:tdde}.

\begin{figure}[!ht] 
\vspace{-.5cm}
	\centering
	\subfigure[\label{fig:isp_info}]{
	    \includegraphics[scale=.33]{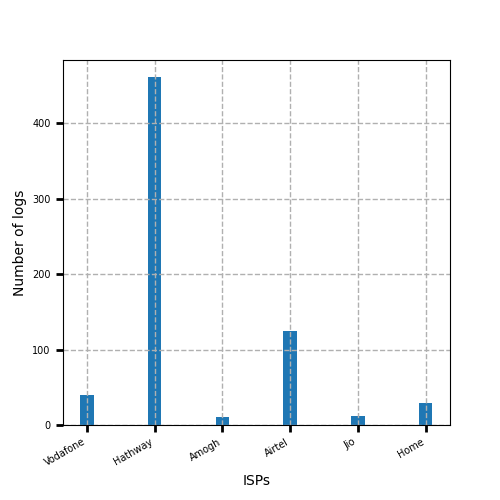}}
	\subfigure[\label{fig:isp_td}]{
		\includegraphics[scale=.33]{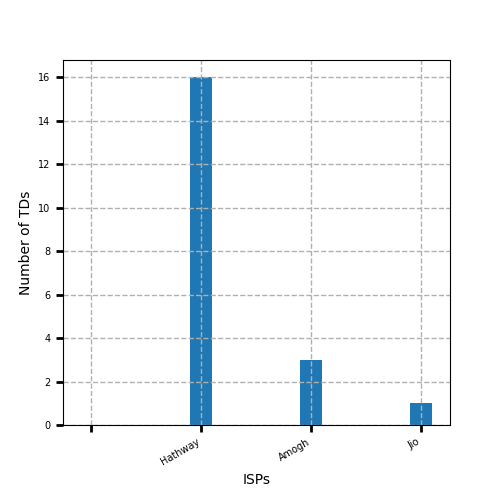}}
	\caption{TD detection summary}
	\label{fig:td_detect_summary}
 \end{figure}

Fig.~\ref{fig:isp_info} shows the ISP-wise breakup of generated logs. There are network logs from $34$ different networks. However, the figure shows networks with a sufficiently large number of network logs only. 

\begin{figure}[!ht] 
\vspace{-.5cm}
	\centering
	\subfigure[\label{fig:nn_violations-hathway}]{
		\includegraphics[scale=.33]{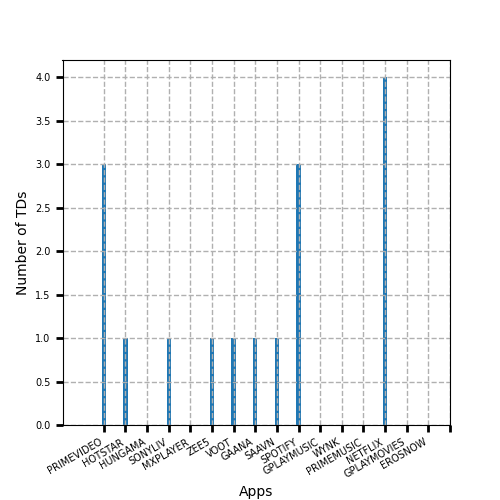}}
	\subfigure[\label{fig:nn_violations-amoghb}]{
		\includegraphics[scale=.33]{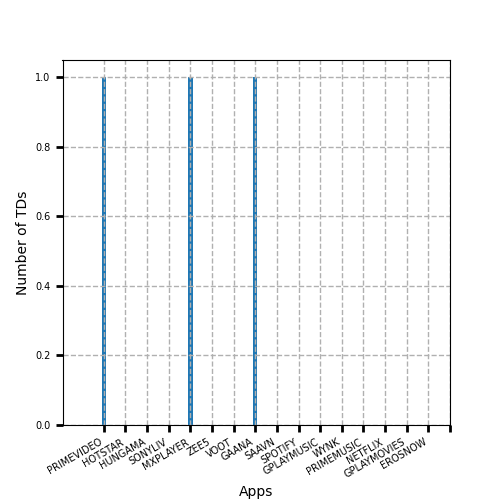}}
	\caption{Detected NN violations: (a) Hathway (b) Amogh Broadband}
	\label{fig:nn_violations_per_isp}
 \end{figure}

\subsection{TD status}
\begin{figure}[htbp]
	\centerline{\includegraphics[width=5cm, height=5cm]{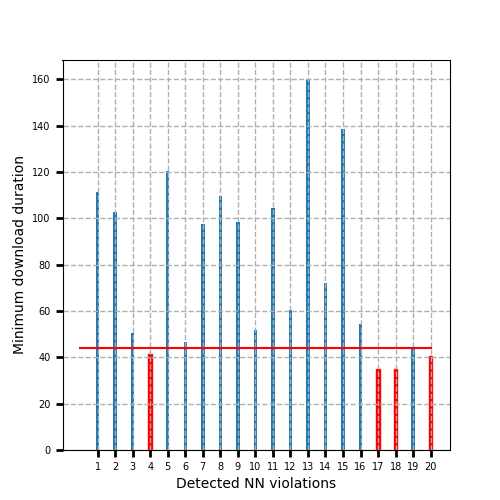}}
	\caption{Variation in minimum data download time in case of NN violation}
	\label{fig:td_test_duration}
\end{figure}

Each log includes performance data for at least two services. The download time for each service varies as per the experienced effect of confounding factors that include deliberate performance degradation. Fig.~\ref{fig:td_test_duration} shows the variation in minimum data download time within each log that has the NN violation (shown by blue and red bars). The red horizontal line indicates the average minimum download time for logs (excluding logs showing a lousy network).  
The logs with minimum download time around average indicate the NN violation in excellent network condition (shown as red bars). These are cases of substantial NN violations due to TD occurring under ample network resource availability. The longer-than-average minimum download time indicates the presence of network congestion. The blue bars indicate the traffic differentiation under such network conditions. The network congestion affects the throughput of all services that increases the possibility of false detection. Our TD detection algorithm's tight thresholds on throughput and the number of low throughput slots ensure correct TD detection.      

Even though traffic management is very prevalent on the Internet, the FairNet has detected net-neutrality (NN) violations only on Hathway, Amogh broadband, and Reliance Jio ISPs only (refer Fig. ~\ref{fig:isp_td} and Fig. ~\ref{fig:nn_violations_per_isp}). This distinction is attributed FairNet's method of comparing similar types of services. Moreover, It shows that per service, NN violations through performance degradation techniques such as traffic shaping are almost absent on the Indian ISPs. The possible explanation is that the technical complexity in practicing such techniques is higher than the benefits. Thus indicating the willingness of ISPs to conform to net-neutrality laws and actively raise concern for any necessary modification in the law \cite{ofcom-ee-complaint}. Note that FairNet detected NN violations for a large number of services on the Hathway network (refer to Fig.~\ref{fig:nn_violations-hathway}). It implies broadband plans for some services that indirectly induce performance degradation in other active services. Hence, resulting in NN violations for affected services.

\section{Conclusion}
\label{sec:conclusion}
We proposed a novel measurement framework named FairNet to detect NN violations of HTTPS traffic over the Internet. Our methodology is based on a client-server architecture where a ``Common Server" (CS) hosts contents from services of interest and replays them on request from the user-client. To test if a service is discriminated by an Internet Services Provider (ISP) in the Internet, FairNet user-client accesses the content of the service under test and content of another service having the similar QoS requirement simultaneously and compares them to check if the service under test is discriminated against the other in terms of service performance. 

The transfer rates at the CS adapted to network conditions using Dynamic Adaptive Streaming over HTTP (DASH) and simultaneously accessing test and reference traffic normalized the congestion and other network confounding factors. Thus our mechanism provided a realistic way of detecting discrimination by comparing services requiring similar quality of service. Our framework uses the appropriate Server Name Indication (SNI) field in the TLS handshake to mimic connection to the original servers of the services. Using a commercial traffic shaper, we established that replayed traffic in our setup is classified as intended traffic from the original services. 

We also developed a novel traffic discrimination detection algorithm that uses time-windowed throughput comparison, socket connection status, and downloaded data size. Through lab experiments and testing with on live network calibrated the parameters of these algorithms. Our framework overcomes several limitations of the existing tools and is easy to deploy and scales to support large number of users. We implemented the FairNet user-client as mobile apps which can be installed for free by users. It enables the end-users to check if their favorite streaming services (both audio and video)  are deliberately throttling by their ISP. Our tool thus acts as a deterrent for any net neutrality violation on the Internet. 

We deployed our apps in the wild and collected traces to detect discrimination. Analysis of traces collected in India indicated that ISPs largely conform to net-neutrality laws as the number of ISPs detected for NN violations are few. The NN violation detection for many services over a single ISP points to the fact that these NN violations are not applied individually on a per-service basis; instead, preferential treatment of some other services over the same network induces such NN violations. Our analysis also shows that the networks tend to rely on larger throughput variations over a much shorter time to manage network resources efficiently.


%

\ifCLASSOPTIONcaptionsoff
  \newpage
\fi



\bibliographystyle{IEEEtran}
\bibliography{IEEEabrv, vinod_bibfile}
\appendices

\section{Customized DASH}
\label{appdx:dash}
The CSS uses a customized dynamic adaptive streaming technique for adaptive modification of data transmission speed. During transmission, a segment is divided into number of bursts that are transmitted with a fixed periodicity, ($TX_{WINDOW}$). Hence the burst size controls the effective transmission speed at the server. The customized adaptive algorithm adapts burst size to match transmission speed with the network conditions. Algo.\ref{algo:dash} shows pseudo-code for the adaptive modification of burst size.  

\begin{algorithm}
	\caption{Customized DASH}\label{algo:dash}
	\begin{algorithmic}[1]
		\State $SP_{min}$ = minimum allowed speed at server
		\State $SP_{max}$ = maximum allowed speed at server
		\State $TX_{WINDOW}$ = burst repetition period
		\State $p_{th} = $throughput at the start of last segment
		\State $c_{th} = $throughput at the end  of the last segment
		\If{$c_{th} > SP_{max}$}
		\State Decrease burst size by $(SP_{max} - c_{th})TX_{WINDOW}$
		\ElsIf{$c_{th} < SP_{min}$}
		\State Increase burst by $(SP_{min} - c_{th})TX_{WINDOW}$ 
		\ElsIf{$c_{th} < p_{th}$}
		\State Decrease burst size by $(pth - c_{th})TX_{WINDOW}$
		\Else
		\State Increase burst size by $(c_{th} - pth)TX_{WINDOW}$
		\EndIf
	\end{algorithmic}
\end{algorithm}

\end{document}